# Machine Learning Application in Online Leading Credit Risk Prediction

Xiaojiao Yu

**Abstract**

Online leading has disrupted the traditional consumer banking sector with more effective loan processing. Risk prediction and monitoring is critical for the success of the business model. Traditional credit score models fall short in applying big data technology in building risk model. In this manuscript, data with various format and size were collected from public website, third-parties and assembled with client's loan application information data. Ensemble machine learning models, random forest model and XGBoost model, were built and trained with the historical transaction data and subsequently tested with separate data. XGBoost model shows higher K-S value, suggesting better classification capability in this task. Top 10 important features from the two models suggest external data such as zhimaScore, multiplatform stacking loans information, and social network information are important factors in predicting loan default probability.

## 1. Introduction

Online leading has gained popularity because of its higher efficiency in offering credit to consumers and small businesses. It is one of the industries that has been disrupted by technology with the electronic lending platform. Loan application decision is made automatically with electronic data driven algorithms [1-2]. Online lenders have the flexibility to offer small loans with short term maturities. Borrowers that are excluded from traditional banking systems can hence have the chance to access credit. Online lending has evolved with platforms connecting lenders and borrowers to more diversified business models such as direct lending, financial institutional partnerships [3].

Online leading bears higher risks compared to traditional bank consumer loans due to insufficient credit checking, inadequate intermediation, lack of transparency and the inherent financial status of typical online borrowers [4]. Therefore, credit risk prediction and management becomes vitally important. Traditional bank loans decision making is based on credit scores with information from application forms and credit reference agency [5]. The objective of the credit score is to evaluate the risk profiles of potential customers and assess their probability of default. It falls in the scope of the discrimination and classification problems [6]. Statistical models and AI models are the most important methods for credit scoring. Statistical models include logistic regression, linear programming, Bayesian models, decision tree and Markov model [7]. The prediction accuracies of the statistical models are usually not high [8]. More recently, artificial intelligence technique such as neural network, support vector machine and nearest neighbor methods are used in the classification task [9]. AI methods do not assume certain data distributions, which are different from statistical models and are superior for nonlinear pattern classification [10]. Ensemble models are more novel techniques that are most recently used in credit risk prediction. First of all, different classifiers are produced and trained with different samples. Classification results are assembled together either by voting or averaging. Bagging, boosting and stacking are often used as ensemble approaches. Nanni and Lumini found that ensemble classifiers can deal with missing data and imbalance classes, performing better classification accuracy [11].

Traditional credit score models are constructed with demographic characteristics, historical payment data, credit bureau data and application data. Credit score and financial history have a strong impact on successful repayment [12]. For online lending, borrower's fraudulent risk is higher. Hence, credit risk models with especially low false negative (type II) errors are desired, since false negative prediction will lead to loss. Other non-standard potential factors that can influence the repayment should be taken into account for model building, training and testing [13]. For instance, soft information such as friendship and photos are reported to have influence in refunding success rate [14]. The collection of credit data has transformed from passive information retrieval to proactive information gathering. The evaluation process of online lending is simpler than traditional process. However, it accesses far more data than a traditional bank in making a loan decision [15].

We used the big data technology by gathering data with different format and size from public website and third-parties and assemble them into a single set of data for each client. Ensemble methods, more specifically, random forest and XGBoost models are developed and trained with historical borrower's information data from our online lending platform. Models are tested subsequently with evaluation metrics such as K-S curve, accuracy, AUC, prevision and recall.

## 2. Data and Variable Definition

The raw data we gathered to build credit risk come from three sources: data from our lending platform, phone records from carriers, third party credit reference companies. Extra feature values are generated from the raw data.

### 2.1 Lending Platform

Our company developed a simple mobile app to enable small short-term loan application, which only requires borrowers' national photo ID, two emergent contacts' name, mobile phone and relationship, and a banking account information. 7 days and 21 days' short-term loan with/without collateral is provided. Borrowing amount can be 1000rmb, 2000rmb for no-collateral loan, 1000rmb up to 6000rmb for loans with collateral.

### 2.2 Phone Records

Phone records data include call records, message, phone bill, nets subflow etc. are collected from the carrier website. Representative data from the carriers' website are listed in table 1. More features can be generated from these raw data.

### 2.3 Third Party Data

Third party credit reference data comes from multiple companies including 91credit, tongdun, zhima credit, qianhaizhengxin. They provide services such as detecting multiple loan stacking or products like credit blacklist database.

Table 1. Representative raw data from different sources.

| App Data : | | | |
|---|---|---|---|
| Income, | categorical | DeviceContact, | categorical |
| Education, | categorical | DeviceCallRecordCount, | numerical |
| MarryStatus, | categorical | MainContact1CallLength, | numerical |
| MainContact1Relation, | categorical | MainContact1CallTimes, | numerical |
| MainContact2Relation, | categorical | MainContact2CallLength, | numerical |
| PositionName, | categorical | MainContact2CallTimes, | numerical |
| JXl_Call_Records Data: | | | |
| contact_count_total, | numerical | contact_min_call_total, | numerical |
| contact_call_count_total, | numerical | contact_min_receive_total | numerical |
| contact_receive_count_total, | numerical | bill_count | numerical |
| contact_min_total | numerical | net_count | numerical |
| Credit Reference Agencies Data: | | | |
| zhima credit data: | | tongdun data: | |
| zhima record count | numerical | td_fraud_medium | numerical |
| Phone_Missing | numerical | td_fraud_high | numerical |
| IP_Missing | numerical | td_no_interloan_blacklist | numerical |
| Bankcard_Missing | numerical | td_no_carrental_blacklist | numerical |
| Phone_Mismatch | numerical | td_no_td_fake | numerical |
| 91 credit data: | | td_no_td_harass | numerical |
| cr91_loan_count | numerical | td_id_criminal | numerical |
| cr91_repayState | numerical | td_multiloan_diff | numerical |
| cr91_borrowState | numerical | td_group_total_count | numerical |
| cr91_borrowAmount | numerical | td_group_multi_app_count | numerical |
| cr91_borrowType | numerical | qianhaizhengxin data: | |
| cr91_loanPeriod_tot | numerical | qhrisk_record_count | numerical |
| cr91_max_loanPeriod | numerical | qhrisk_source_id_count | numerical |
| cr91_min_loanPeriod | numerical | qhrisk_risk_score_total | numerical |
| cr91_arreasAmount_total | numerical | qhrisk_risk_mark_count | numerical |

3. Models developed

Ensemble methods, random forest and XGBoost are developed to predict probability of default of potential clients with overall 8990 features (both from raw data and generated data). Categorical features were converted to numerical values with label encoder from scikit-learn for model training purpose. The data were partitioned into two subsets: 70% of the data were used to train models, 30% were used for testing. Training data and test data were randomly selected. Overall 211357 observations were used for modeling training. 90462 observations were used for testing. Observations with missing values were not taken into account.

3.1. Random forest model parameter tuning

Random forest builds a collection of decision trees on bootstrapped training samples. It reduces the correlations of the decision trees in the forest by randomly choosing splitting attributes from the full attributes when constructing each decision tree.

Thus, it is superior to the traditional bagged trees[16]. Predictions from each tree is aggregated. Classification is assigned with the majority votes. Important hyper parameters for random forest model are number of trees (no_trees), sample_split, sample_leaf. They define the number of decision trees we use to build the random forest. Sample_split defines the minimum number of samples required to split an internal node. Sample_leaf defines the minimum number of samples required at a leaf node. To set the optimal value of these hyper parameters, precision recall curve ( precision = $\frac{tp}{tp+fp}$ , recall=$\frac{tp}{tp+fn}$) was plotted for different hyper parameter values when trained model was applied on the testing samples.

Clearly, PR curve shifts to the right when the number of decision trees increases from 50 to 500, suggesting higher precision scores when recall value is the same and higher recall value when precision score is the same. Further increase of the number of decision trees to 5000 from 500 does not shift the PR curve to the right noticeably as shown in Fig. 1. Different values of sample_split do not seem to affect the precision and recall value observably, Fig. 2. When sample_leaf increases from 1 to 3 and 5, PR curve shifts to the right slightly, Fig. 3. By experimenting with the hyper parameter values, no_trees was set to be 5000, sample_split was set at 2, and sample_leaf was set at 1.

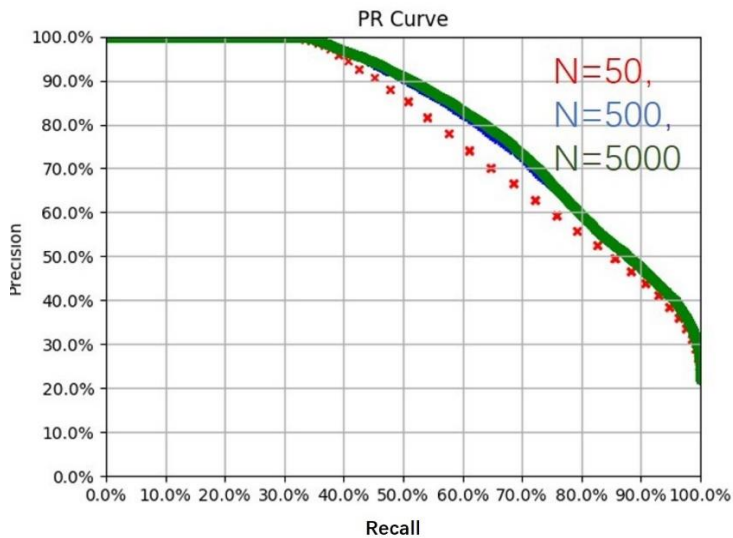

Fig 1. Precision-Recall curve of random forest model for different number of trees: N=50, 500, 5000.

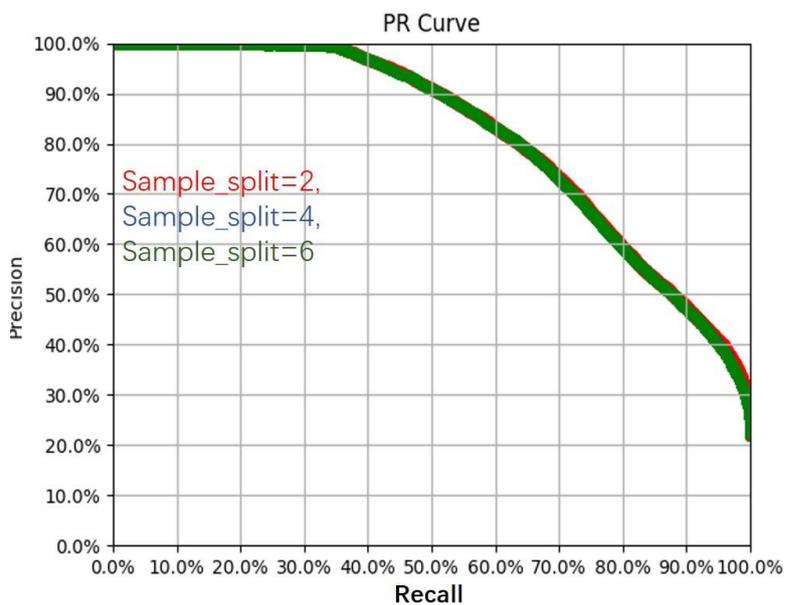

Fig 2. Precision-Recall curve of random forest model for sample_split value to be 2, 4, 6.

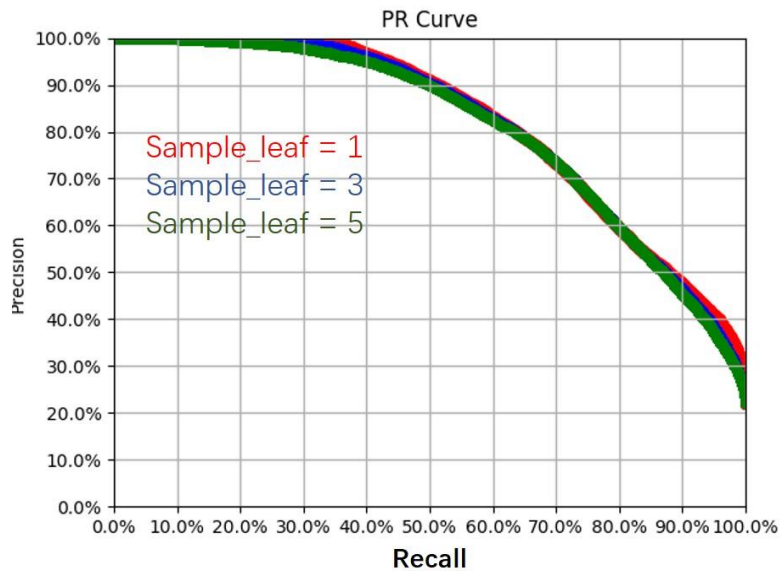

Fig 3. Precision-Recall curve of random forest model for sample_leaf value to be 1, 3, 5.

**3.2 XGBoost fine parameter tuning**

XGBoost belongs to the Gradient Boosting algorithm family. It is implemented with c++ and provides parallel tree boosting that can give more accurate solutions faster compared to existing solutions. XGBoost model is trained in an additive manner of all the decision tree predictors. For each tree, subsampling is possible over objects and features to prevent overfitting. Optimization is performed on the prior prediction residuals. Each tree is trained after all the previous trees' structures are optimized. At each step, objective function includes a loss function and a regularize item to prevent overfitting. Tayler expansion of the loss function is taken up to the second order [17]. XGBoost allows users to define customized objective functions.

XGBoost algorithm has a lot of hyper parameters for tuning. Among them, max_depth (the maximum depth of a tree), eta (Learning rate), colsample_bytree (the fraction of columns to be randomly samples for each tree), subsample (the fraction of observations to be randomly samples for each tree), min_child_weight (minimum sum of weights of all observations required in a child), alpha (L1 regularization term on weight) and gamma (Gamma specifies the minimum loss reduction required to make a split) are the most important ones. Optimal values of these parameters were set by evaluating the AUC (area under the roc curve) value when trained model was used to predict testing samples. The higher value of AUC curve, the better of the model in predicting testing samples.

Varying the value of max_depth and keeping other hyper parameters the same, it is seen that when max_depth is 20, AUC of the testing set ranks the highest among all the tested values. Similarly, optimal values of other hyper parameters were set as in table 2. It is seen that when learning rate is 0.01, Colsample_bytree value is 0.7, subsample is 0.7, Min_child_weight is 1, gamma is 0.01, alpha is 0.1, validation set has the highest AUC value, which is 0.94064.

Table 2 Hyper parameters tuning for XGBoost model

| Max_depth | eta | Colsample_bytree | subsample | Min_child_weight | gamma | alpha | Val_auc |
|---|---|---|---|---|---|---|---|
| 20 | 0.01 | 0.7 | 0.7 | 1 | 0.01 | 1 | 0.938943 |
| 10 | 0.01 | 0.7 | 0.7 | 1 | 0.01 | 1 | 0.919428 |
| 5 | 0.01 | 0.7 | 0.7 | 1 | 0.01 | 1 | 0.853687 |
| 20 | 0.1 | 0.7 | 0.7 | 1 | 0.01 | 1 | 0.937884 |
| 20 | 0.3 | 0.7 | 0.7 | 1 | 0.01 | 1 | 0.930698 |
| 20 | 0.01 | 0.8 | 0.8 | 1 | 0.01 | 1 | 0.939045 |
| 20 | 0.01 | 0.6 | 0.6 | 1 | 0.01 | 1 | 0.938668 |
| 20 | 0.01 | 0.7 | 0.7 | 3 | 0.01 | 1 | 0.936187 |
| 20 | 0.01 | 0.7 | 0.7 | 5 | 0.01 | 1 | 0.934721 |
| 20 | 0.01 | 0.7 | 0.7 | 1 | 0.1 | 1 | 0.938715 |
| 20 | 0.01 | 0.7 | 0.7 | 1 | 0.01 | 0.1 | 0.94064 |
| 20 | 0.01 | 0.8 | 0.8 | 1 | 0.01 | 0.1 | 0.940533 |

4. Experimental results

Top 10 important features from both the random forest model and the XGBoost model were plotted after the training process. Hence, we can identify the most important factors in determining whether a loan should be lent to a potential customer. Among the 8990 total features, zhimaScore, td_multi_platform_6mon_1888318_cnt (tongdun multi-platform stacking loan in the past 6 months count), td_multi_platform_1mon_1888314_perc (tongdun multi-platform loan in the past 1-month count percentage), td_multi_platform_12mon_1888320_cnt (tongdun multi-platform loan in the last 12 months count), Name_Match_Reliability_Good, td_multi_platform_7d_1888312_perc, Name_Match_Sharing_Good, td_multi_platform_60mon_1888326_cnt, deviceContactCount, td_multi_platform_24mon_1888324_cnt. The percentage of these top 10 important features are marked in Fig.4.

Top 10 important features from the XGBoost model are zhimaScore, td_multi_platform_1_mon_1888314_perc, deviceContactCount, td_multi_platform_7d_1888312_perc, td_multi_platform_3mon_1888316_cnt, td_multi_platform_6mon_1888318_cnt, credooScore, deviceCallRecordCount, contact_no_rece_cnt_perc_6_mon, td_multi_platform_12mon_1888320_cnt.

Top ten important features from both models are almost the same except for their rankings. An obvious difference is that zhimaScore ranks especially high in random forest model, which is 3.67%, the other 9 features are all about 0.16%. Percentage of top 10 important features in XGBoost model does not vary that much compared to random forest model.

Since both the random forest model and XGBoost model involves random feature sampling and observation sampling when each decision tree is built, the resulted top 10 important features can vary accordingly. The correlation of the decision trees in the random forest model depends on the random observations and feature sampling, which determines the effectiveness of the random forest model in classification. For random forest model, max. features for building each decision tree is set at the default value, which is the square root of 8990. This is substantially smaller than the colsample_bytree value in XGBoost model. Also in the XGBoost model, regularization is added in optimizing the objective function to prevent overfitting. This might be the reason XGBoost model outperforms random forest model.

Lorenz curves from both models are plotted in Fig. 5. They show the cumulative percentage of bad cases and good cases with the percentage of samples. K-S value is defined as the difference between cumulative percentage of bad cases and good cases. K-S values from both models are calculated from the curve. The maximum value of K-S from the random forest model is 0.6474, from XGBoost is 0.7203. Since K-S measures the distance between the cumulative percentage of bad cases and good cases, the higher of K-S value, the better of the model in separating bad and good cases. By looking at the trained model size, XGBoost model also has smaller model size compared to random forest model. Clearly, XGBoost model outperform random forest model in this classification task.

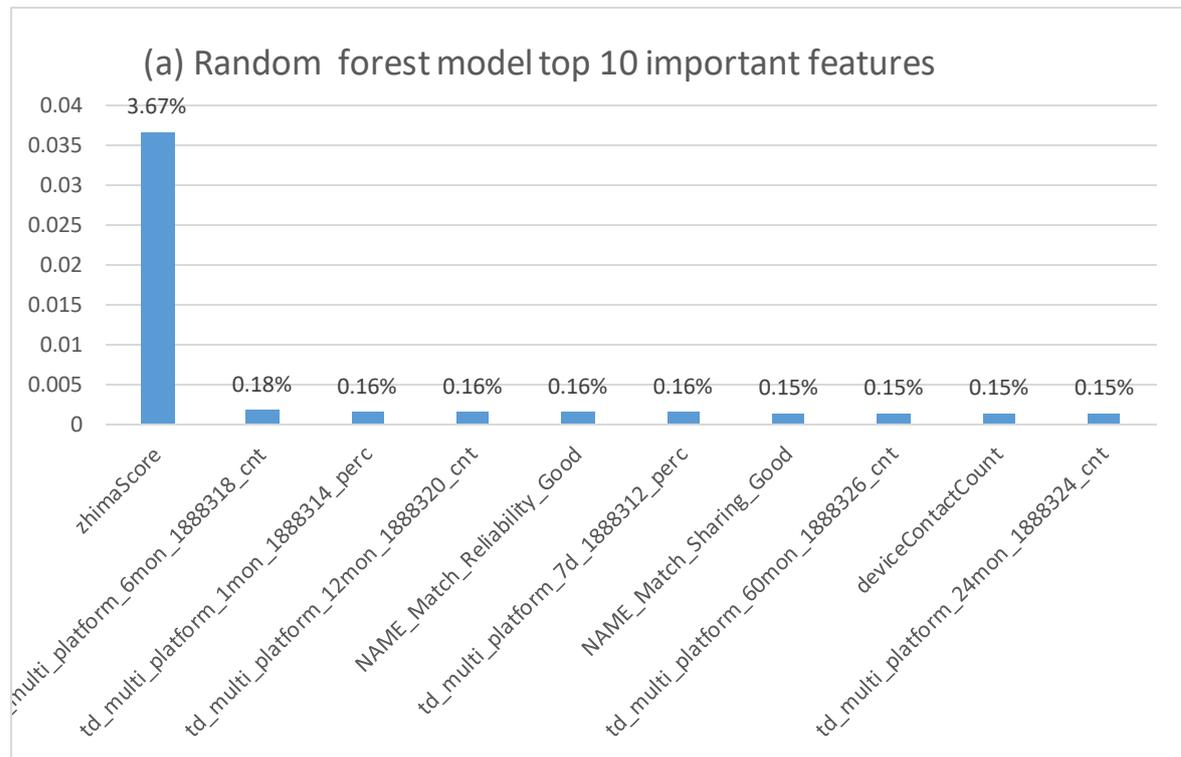

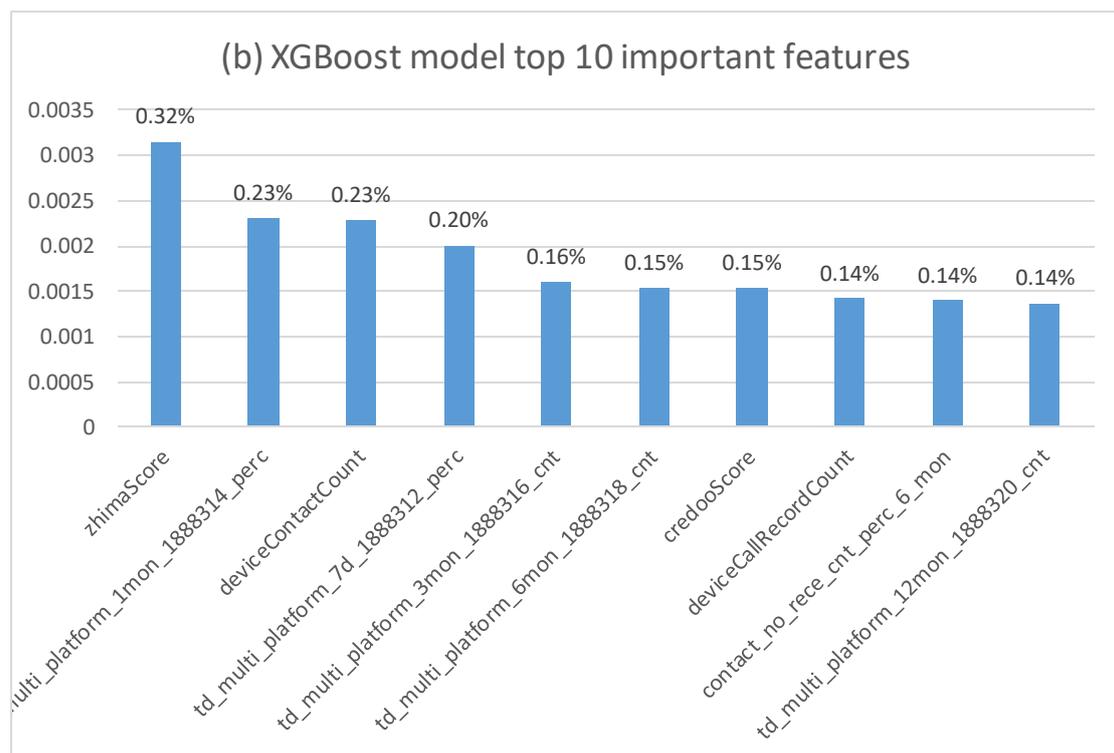

Fig. 4. Top 10 important features from (a) Random forest model. (b) XGBoost model

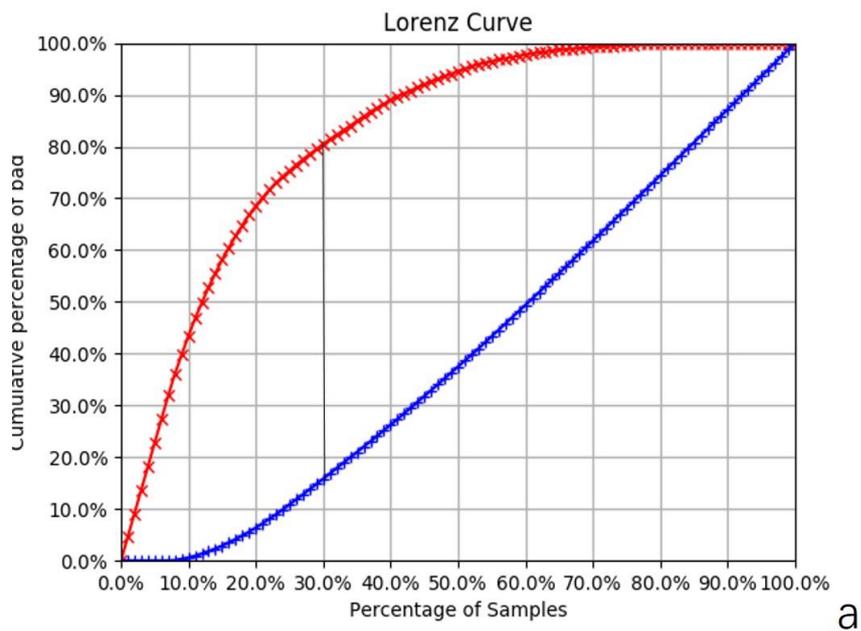

a

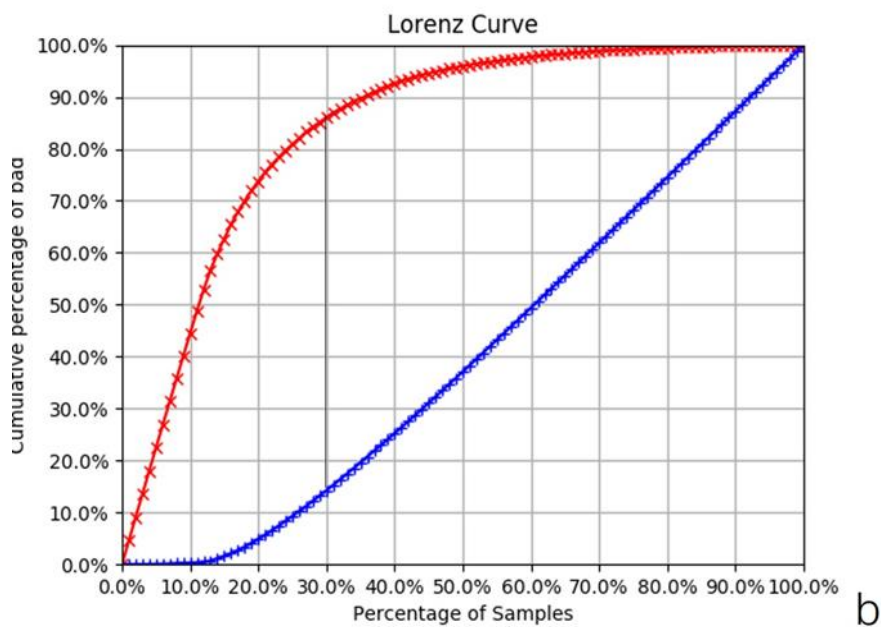

b

Fig 5 (a). K-S curve for Random forest model. (b). K-S curve for XGBoost model

### 5. Conclusion

In this manuscript, a random forest model and XGBoost model were trained with the combination of internal app data and external third party credit reference and carrier data to predict credit default probability of potential borrowers. Each model was optimized by tuning hyper parameters and evaluated with testing data. Top 10 important features of the two models were plotted and ranked. XGBoost model shows homogeneous features importance percentage compared to random forest models. Both of the two models show that external data such as zhimaScore, multiplatform stacking loans information, and social network information are important factors in predicting loan default probability. K-S value from both models suggests better classification of bad cases and good cases of XGBoost.